\documentclass[twocolumn,amsmath,amssymb,superscriptaddress]{revtex4}

\usepackage{graphicx}
\usepackage{multirow}

\usepackage[breaklinks]{hyperref}
\hypersetup{
    bookmarks=false,
    pdfstartview={FitH},
    colorlinks=true,
    linkcolor=blue,
    citecolor=blue,
    urlcolor=blue
}

\begin{document}

\title{Resonant soft x-ray emission as a bulk probe of correlated electron
behavior in metallic Sr$_x$Ca$_{1-x}$VO$_3$}

\author{J.~Laverock}
\author{B.~Chen}
\author{K.~E.~Smith}
\affiliation{Department of Physics, Boston University, 590 Commonwealth Avenue,
Boston, Massachusetts, MA 02215, USA}

\author{R.~P.~Singh\footnote{Present address:
Department of Physics, IISER Bhopal, MP-462023, India}}
\author{G.~Balakrishnan}
\affiliation{Department of Physics, University of Warwick, Coventry, CV4 7AL,
United Kingdom}

\author{M.~Gu}
\affiliation{Department of Physics, University of Virginia,
Charlottesville, VA 22904, USA}
\author{J.~W.~Lu}
\affiliation{Department of Materials Science and Engineering,
University of Virginia, Charlottesville, VA 22904, USA}
\author{S.~A.~Wolf}
\affiliation{Department of Physics, University of Virginia,
Charlottesville, VA 22904, USA}
\affiliation{Department of Materials Science and Engineering,
University of Virginia, Charlottesville, VA 22904, USA}

\author{R.~M.~Qiao}
\author{W.~Yang}
\affiliation{Advanced Light Source, Lawrence Berkeley National Laboratory,
Berkeley, California, CA 94720, USA}

\author{J.~Adell}
\affiliation{MAX-lab, Lund University, SE-221 00 Lund, Sweden}

\begin{abstract}
The evolution of electron correlation in Sr$_{x}$Ca$_{1-x}$VO$_3$ has been
studied using a combination of
bulk-sensitive resonant soft x-ray emission spectroscopy (RXES),
surface-sensitive photoemission spectroscopy
(PES), and {\em ab initio} band structure calculations.
We show
that the effect of electron correlation is enhanced at the surface. Strong
incoherent Hubbard subbands are found to lie $\sim 20$\% closer in energy
to the coherent
quasiparticle features in surface-sensitive PES measurements
compared with those from bulk-sensitive RXES,
and a $\sim 10$\% narrowing of the overall bandwidth at the surface is also
observed.
\end{abstract}

\maketitle
Understanding correlated electron behavior remains one of the most
important problems in condensed matter physics. In correlated
electron systems, the interaction between electrons is of the order of, or
larger than, the electron kinetic energy, and the concept of a well-defined
quasiparticle is restricted to a narrow region of energies near the Fermi
level, beyond which our strict understanding of a quasiparticle with a
defined dispersion relation, easily accessible through band theory, breaks
down. Rather, the competition between itinerant delocalized behavior and
strongly-correlated localized behavior leads to the transfer of spectral
weight away from the quasiparticle peak (QP) and into incoherent Hubbard
subbands.\cite{kotliar2006}

As a probe of correlated electron behavior, angle-resolved photoemission
spectroscopy (ARPES) is uniquely placed: It is capable of measuring the
many-body electron dispersion relation, $E({\bf k})$, with high precision,
including the Fermi surface (FS) and modifications to $E({\bf k})$
that are due to the coupling of the electrons to collective
excitations.\cite{kevan1991etc}
However,
owing to the short mean free path of the photoelectrons, ARPES is intrinsically
a surface sensitive probe, particularly of three-dimensional (3D) systems,
and is usually limited to within 15~{\AA} of the surface (although efforts
to enhance the bulk sensitivity, such as laser-ARPES
can yield information to over 100~{\AA}).\cite{koralek2006}
On the other hand, resonant soft
x-ray emission spectroscopy (RXES), being a photon-in--photon-out technique,
is truly a probe of the {\em bulk} electronic structure, with a typical
sampling depth of $\sim 100$~nm.\cite{smith2003}

Here, we demonstrate that RXES is a sensitive probe of correlated
electron behavior, capable of yielding complementary information to
ARPES from a truly {\em bulk} perspective.
Specifically, we show that electron-electron correlation in
Sr$_x$Ca$_{1-x}$VO$_3$ is enhanced near the sample surface.
We accomplish this through comparison of RXES measurements with
photoemission spectroscopy (PES)
measurements on the same Sr$_x$Ca$_{1-x}$VO$_3$ single crystal samples,
as well as on CaVO$_3$ thin film samples.
Our RXES results are in much
closer agreement with dynamical mean-field theory (DMFT) calculations than the
PES results; in these calculations
the value of the Hubbard $U$ parameter has been determined
from first principles.\cite{nekrasov2005} Our RXES data explain the
discrepancy between numerous published ARPES
measurements of Sr$_x$Ca$_{1-x}$VO$_3$,%
\cite{sekiyama2004,eguchi2006,yoshida2005etc,yoshida2010,aizaki2012}
and between ARPES and DMFT calculations.%
\cite{nekrasov2005,nekrasov2006etc,ishida2006} Moreover, these results
illustrate the powerful application that RXES
can have in addressing correlated electron behavior,
particularly when comparisons can be directly made with (AR)PES measurements,
and they shed light on the different scattering mechanisms of the two
techniques.

The Sr$_x$Ca$_{1-x}$VO$_3$ ($0 \leq x \leq 1$) family of oxides
are prototypical strongly-correlated materials,
exhibiting both strong Hubbard subbands as well as appreciable quasiparticle
peaks.\cite{nekrasov2005,yoshida2010} These materials have been well
studied with ARPES, but the results have yielded conflicting pictures of the
role of electron correlations.  Measurements aimed at extracting ``bulk''
ARPES spectra have been performed at both high\cite{sekiyama2004} and
low\cite{eguchi2006} incident photon energies (i.e.~away from the minimum in the
photoelectron mean free path). These studies reported either spectra that were
independent of $x$,\cite{sekiyama2004} or a weak suppression of the QP in
CaVO$_3$.\cite{eguchi2006} More recent ARPES measurements, in which FSs and
band dispersions were clearly observed in both end-members, suggest both a
narrowing of the bandwidth and a suppression of the QP in CaVO$_3$ compared
with SrVO$_3$,\cite{yoshida2010} consistent with expectations that CaVO$_3$
experiences stronger electron correlations. The role of the surface in ARPES
measurements has been questioned throughout,\cite{liebsch2003,ishida2006}
and confirmation from a truly bulk probe is clearly desirable.

In the transition-metal $L$-edge RXES process, a $2p$ core electron is
excited into the conduction band in analogy with the x-ray absorption
process.  The excited state subsequently decays to fill the core hole,
in which the energy and, in principle momentum, of the emitted x-ray is
measured.\cite{kotani2001etc} The core hole can decay via several
routes: (i)~elastic scattering, in which the excited electron returns to the
original core level
without transferring energy to the system, (ii)~inelastic scattering, in
which energy is transferred to the system during the intermediate state,
in the form of either localized or delocalized excitations (i.e.~Raman-type
resonant inelastic x-ray scattering, RIXS), and (iii)~fluorescent-like core hole
decay,
in which an electron from an occupied valence band state makes the transition to
the core hole. Both (i) and (ii) yield
features whose energy depends on the incident photon, whereas the energy of
(iii) is independent of the excitation energy. In this work, we will focus
on (iii), i.e.~the normal x-ray emission--like part of the spectrum.

Large high quality single crystals of CaVO$_3$ (CVO) and
Sr$_{0.5}$Ca$_{0.5}$VO$_3$ (SCVO) were grown by the floating zone technique
in a four mirror optical furnace, employing growth rates of 7 to 10~mm/h in
an atmosphere of 1~bar of Ar + 3\% H$_2$ gas.\cite{inoue1998}
Samples for the RXES measurements
were obtained by cleaving the as grown crystals {\em ex-situ},
and were immediately loaded into the ultra-high
vacuum chamber. RXES and x-ray absorption spectroscopy (XAS) measurements were
performed at beamline 8.0.1 of the Advanced Light Source, Lawrence Berkeley
National Laboratory. The resolution of the emission spectrometer was 0.35~eV
at full-width half maximum (FWHM).  XAS measurements, recorded in both total
electron yield (TEY, sampling depth $\sim$~10~nm) and total fluorescent yield
(TFY, sampling depth $\sim 100$~nm) were recorded with an energy resolution
of 0.2~eV. For ARPES measurements, clean well-ordered surfaces were prepared
by cleaving {\em in-situ} in ultra high vacuum,
and were oriented with reference to low-energy
electron diffraction patterns. (AR)PES measurements were performed at
beamline I4
at the MAXlab synhrotron radiation facility (Lund, Sweden).
The incident photon
energy was 80~eV and the total instrument electron energy resolution was
25~meV. Measurements
of the FS and band dispersion of both samples (not presented here) are in
good agreement with previously published work.\cite{yoshida2010,aizaki2012}
The sampling depth of the RXES measurements is estimated to be $\sim 100$~nm,
which can be compared with the photoelectron mean free path of $\sim
0.5$~nm for our ARPES measurements (at 80~eV). Correspondingly, previous
``bulk-sensitive'' ARPES measurements recorded at a photon energy of 900~eV
reach less than 2~nm into the sample (in that work, the surface
depth was estimated to be 0.5~--~1~nm).\cite{sekiyama2004}
Thin (46~nm) CVO films were grown on
SrTiO$_3$(001) substrates using a pulsed electron-beam deposition
technique,\cite{gu2013}
and were capped with a 2.5~nm SrTiO$_3$ layer to protect
the surface of CVO, precluding measurements at the O $K$-edge and ARPES.
{\em Ab initio} calculations of the electronic structure of
cubic perovskite SrVO$_3$ and orthorhombic (distorted
perovskite)\cite{falcon2004}
CaVO$_3$ were performed with the all electron full
potential linearized augmented plane wave {\sc Elk} code\cite{elk}
within the local density approximation (LDA).  Convergence was achieved on
84 k-points in the irreducible Brillouin zone (IBZ) of cubic SrVO$_3$ and
on 343 k-points in the orthorhombic IBZ of CaVO$_3$.

\begin{figure}[t!]
\begin{center}
\includegraphics[width=0.90\linewidth,clip]{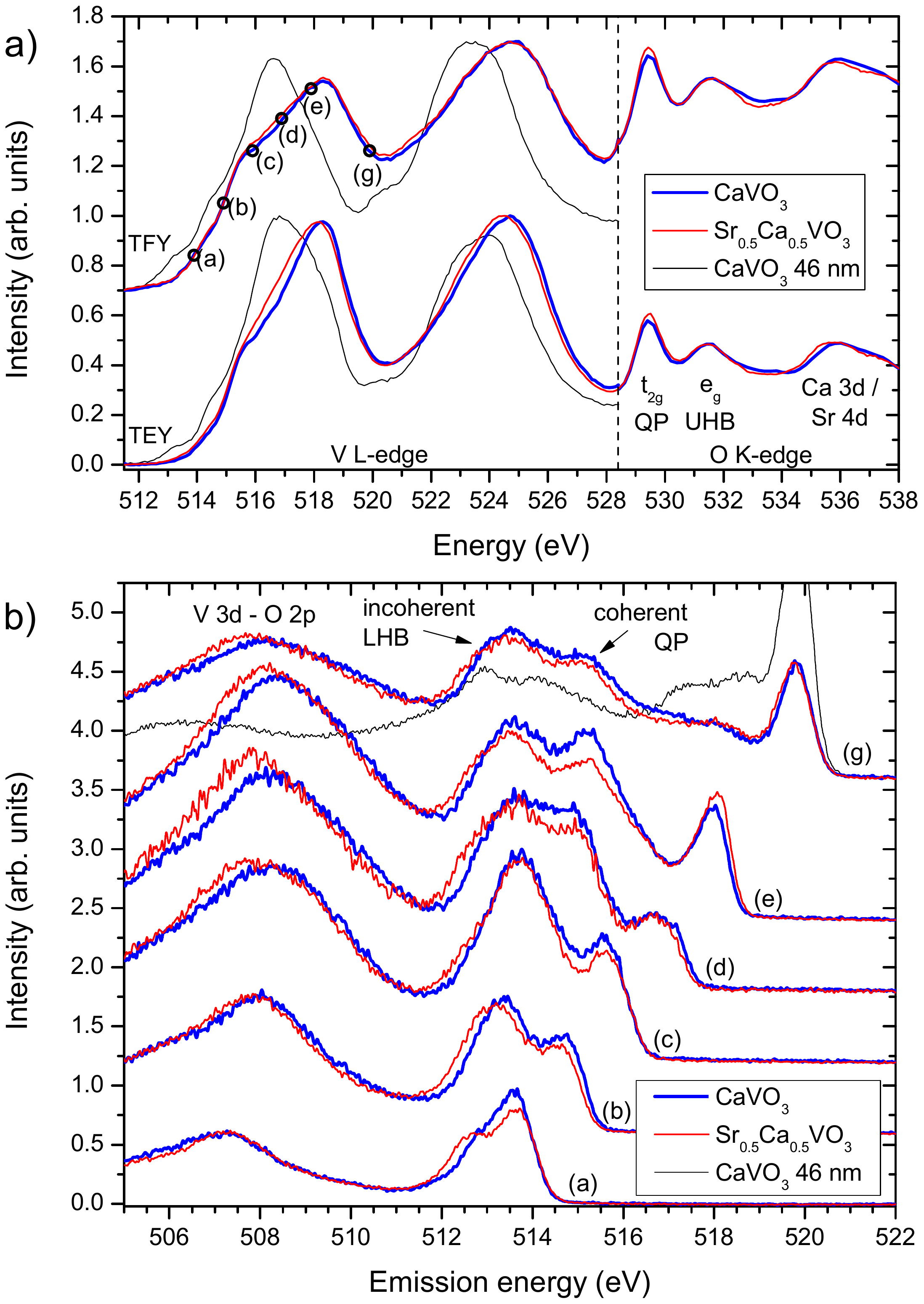}
\end{center}
\vspace*{-0.2in}
\caption{\label{f:rxes} (Color online) Bulk (a) XAS and (b)
RXES measurements of CVO and SCVO.
The spectra of a 46~nm thin film of CVO\cite{gu2013} are also shown
for comparison. The excitation energies used in the RXES measurements of
(b) are shown by the circles in (a). The vertical line in (a) indicates
separate V $L$-edge (left) and O $K$-edge (right) measurements, which have been
separately scaled for clarity.}
\end{figure}

V $L$-edge and O $K$-edge
XAS and RXES spectra are shown for CVO and SCVO in Fig.~\ref{f:rxes}. XAS
spectra at the V $L$-edge are very similar for the two compounds (shown in
both TEY and TFY modes in Fig.~\ref{f:rxes}a), and are in close agreement
with previous measurements.\cite{pen1999} These spectra are not well
described by atomic multiplet calculations owing to the itinerant behavior
of the metallic V $d$ electrons.\cite{pen1999} At the O $K$-edge, which is
sensitive to the unoccupied conduction band
partial density of states (PDOS), three features
dominate the spectra and are related to excitations into V $t_{2g}$ and $e_g$
states, as well as excitations into the Ca/Sr $d$ states at higher energies.
DMFT calculations of the V $t_{2g}$ states predict that the upper Hubbard band
(UHB) overlaps with the $e_g$ feature, whereas the $t_{2g}$ XAS feature
encompasses the coherent QP peak.\cite{nekrasov2005} An alternative
description, based on extended cluster model calculations, also suggests
that the $t_{2g}$ feature involves excitations into the coherent metallic
states.\cite{mossanek2010} In Fig.~\ref{f:rxes}a, the ratio of the $t_{2g}$
to $e_g$ intensities is found to be slightly higher in SCVO than in CVO,
reflecting the larger QP weight that accompanies the ``weaker'' correlations
in this compound, in agreement with DMFT\cite{nekrasov2005} as well as
ARPES measurements of the occupied states.\cite{yoshida2010}

RXES measurements across the V $L_3$-edge are shown in Fig.~\ref{f:rxes}b. At
low energies, the peak centered at 508~eV is due to V $3d$ -- O $2p$
hybridization,\cite{laverock2012} whereas the V $3d$ states appear in the
range 512~--~516~eV. The constant emission energy of this feature establishes
its origin as fluorescent-like decay of the core hole rather than Raman-type
loss features, which are weak for metallic systems.  The dispersive peak is
predominantly due to elastically scattered x-rays, and is centered at the
incident photon energy of each spectrum. As the incident photon energy is
tuned through the V $L_3$-edge, emission from the available V $3d$ states
is resonantly enhanced. For spectrum (f), which is recorded above the V
$L_3$-edge absorption feature, the resonance effects on the fluorescent part
of the spectrum are weak, and we interpret this spectrum as most closely
representing the occupied V $3d$ PDOS.  Moreover, at this energy, low-energy
loss features (which are typically $\lesssim 4$~eV)\cite{laverock2012}
are well separated from
the fluorescent part of the spectrum.

In spectra (d-f), the V $3d$ fluorescence is split into a double-peaked
structure, which can be associated with the incoherent lower Hubbard
band (LHB) and coherent QP electron states, as suggested by LDA+DMFT
calculations\cite{nekrasov2005} and observed in ARPES
measurements.\cite{yoshida2010}
Whilst the energy of the coherent feature is similar
for both compounds, the incoherent feature of the Sr-doped compound is
spread to lower energies, and the peak is centered 0.1~--~0.2~eV lower
than CVO. Quantitatively, we find that the separation between coherent and
incoherent features is 1.4 and 1.6~eV for CVO and SCVO respectively. Note
that this is the separation between the centers of the
two RXES features, and does not necessarily reflect the fundamental energy
separation between LHB and QP states. Nevertheless, the evolution in this
separation can be associated with the evolution in electron correlations
in going
from SrVO$_3$ to CaVO$_3$, usually identified as a consequence of the narrowing
of the bandwidth.  The bandwidth is difficult to experimentally assess from
these RXES spectra, owing to the broad nature of the features. However,
by analyzing the first derivative of the spectra, it is possible to
estimate the total bandwidth of the incoherent + coherent V $3d$ states,
which we find to be 3.0 and 3.2~eV for CVO and SCVO respectively (i.e.~an
increase of $\sim 10$\%, consistent with our band structure calculations).
In our surface-sensitive ARPES spectra, the bottom of the $t_{2g}$ band (at
$\Gamma$) was observed at 0.44~eV and 0.49~eV for CVO and SCVO respectively,
in agreement with these bulk estimations.

\begin{figure}[t!]
\begin{center}
\includegraphics[width=0.90\linewidth,clip]{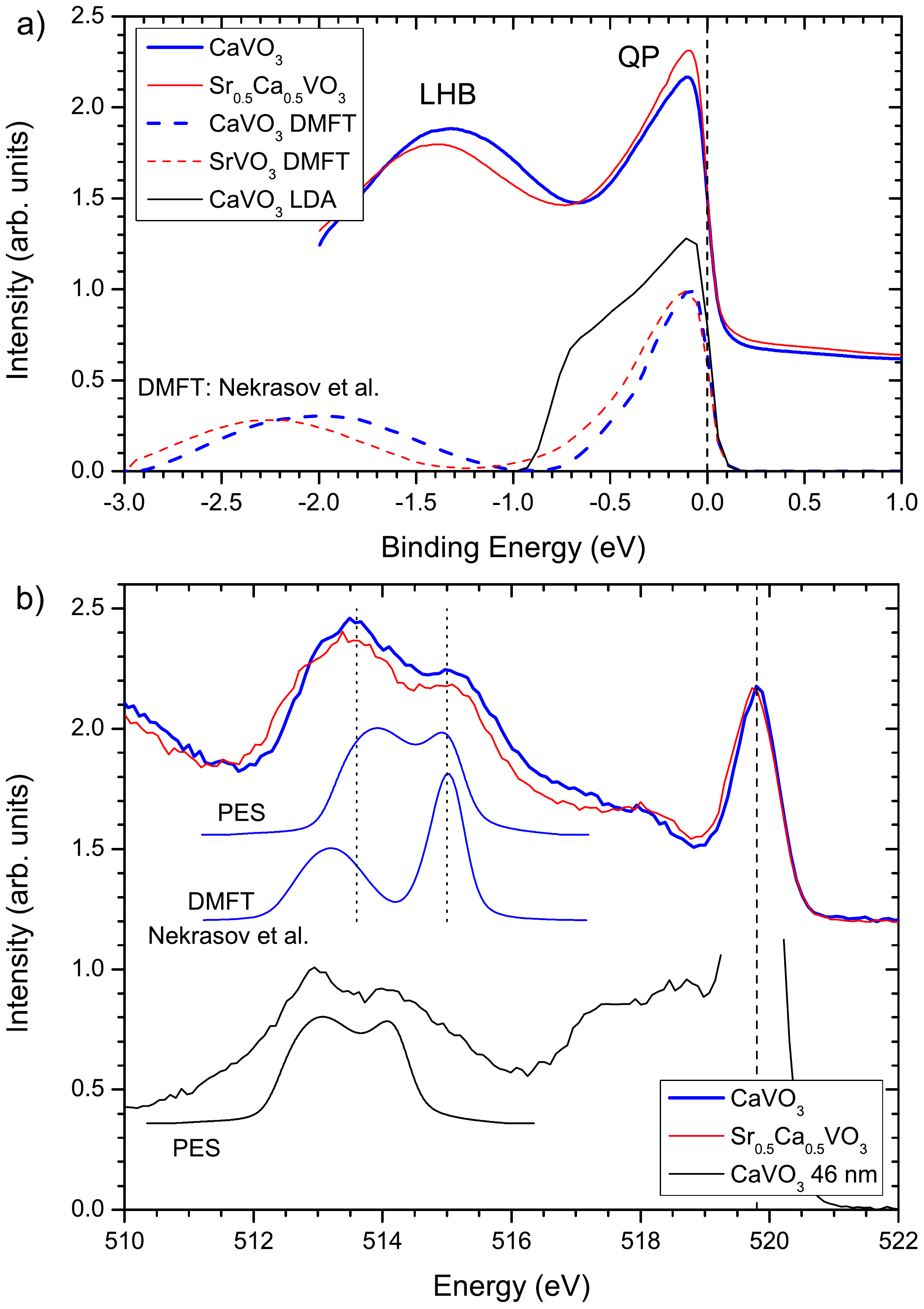}
\end{center}
\vspace*{-0.2in}
\caption{\label{f:rixs} (Color online) Comparison of
PES
and RXES spectra. (a) PES spectra of CVO and
SCVO compared with the LDA DOS of CaVO$_3$ and DMFT
calculations of CaVO$_3$ and SrVO$_3$ (reproduced from
Ref.~\cite{nekrasov2005}). (b) PES spectra compared with the RXES spectra at
excitation energy (f). The PES and DMFT spectra have been broadened to
approximate the RXES measurement (see text), and the RXES spectra are shown
rebinned to double the statistical precision.}
\end{figure}

Additional support for this interpretation is provided through comparison
with thin film (46~nm) CaVO$_3$. Spectra of this sample at the V $L$-edge
are shown alongside the single crystal data in Fig.~\ref{f:rxes}. In the
XAS, the thin film spectra are markedly different from those of the bulk
single crystals, and more closely resemble the results of atomic multiplet
calculations of the V $d^1$ ion\cite{pen1999} and other V$^{4+}$ oxides
(e.g.~VO$_2$)\cite{laverock2012}, indicating the suppression of itinerant
electron behavior and the corresponding relevance of the localized electron
picture. The strong elastic peak and relative enhancement of Raman-type
loss features in the RXES (Fig.~\ref{f:rxes}b) are also indicative of more
localized electron behavior. Moreover, the separation between the incoherent
and coherent V $3d$ features is much closer than for the single crystal
CaVO$_3$ at 1.1~eV, $\sim 20$\% lower than bulk CaVO$_3$.  Correspondingly,
the overall (coherent + incoherent) bandwidth of thin film CaVO$_3$ is
reduced by $\sim 10$\% to 2.7~eV.  Together, these results indicate that the
electron correlations in thin film CaVO$_3$ are substantially exaggerated
compared with the bulk material. Indeed, for thicknesses of $\lesssim
4$~nm, CaVO$_3$ has been found to exhibit a metal-insulator transition
at $\sim 100$~K.\cite{gu2013} RXES measurements of 4~nm thick CaVO$_3$
(not presented here owing to the weak count rate of these very thin samples)
showed similar spectral features to 46~nm samples.

\begin{figure}[t!]
\begin{center}
\includegraphics[width=0.90\linewidth,clip]{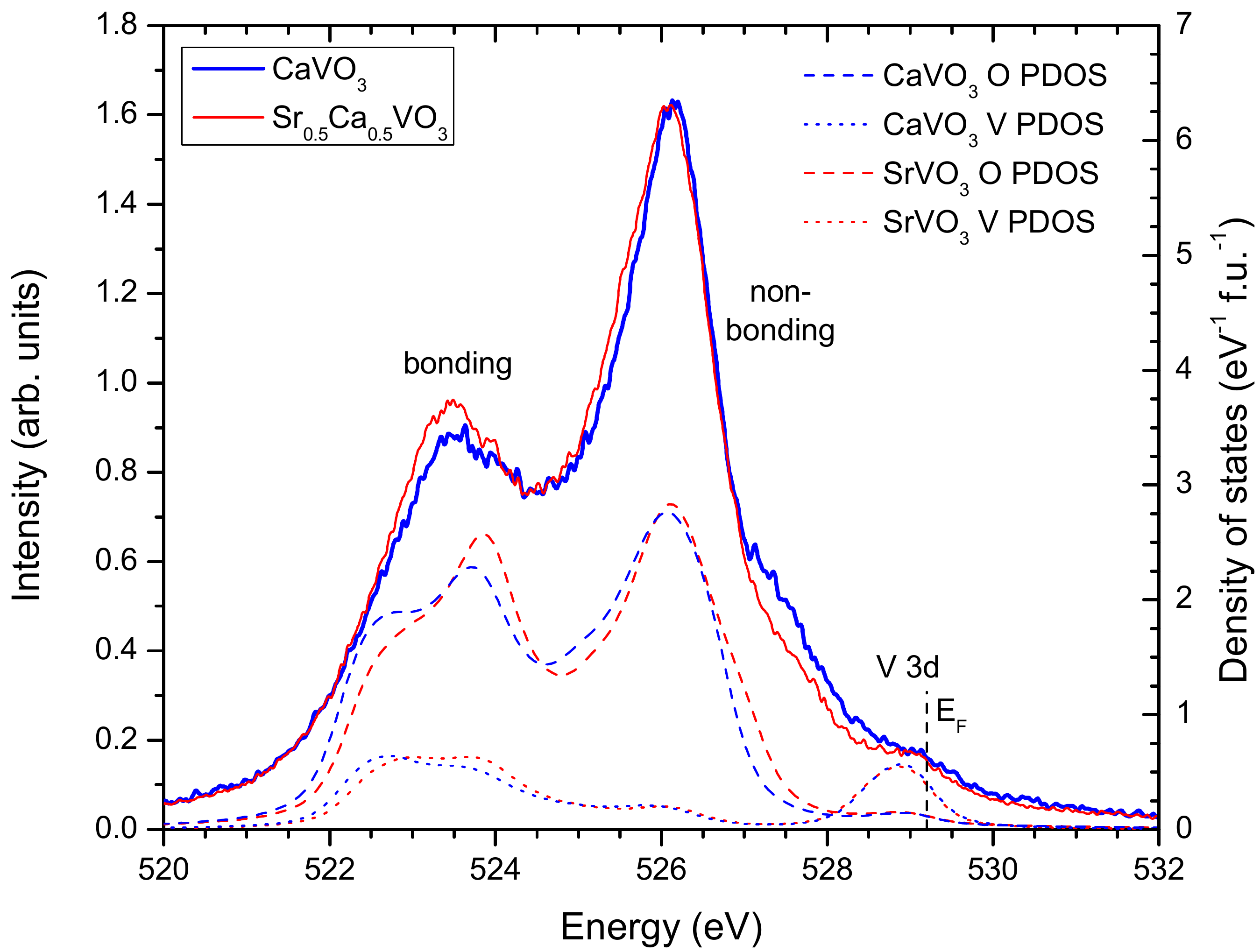}
\end{center}
\vspace*{-0.2in}
\caption{\label{f:okxes} (Color online) O $K$-edge XES spectra of CVO and
SCVO. The corresponding O and V PDOS from the
LDA are shown for comparison after convolution with a Lorentzian of 0.22~eV
HWHM and a Gaussian of 0.35~eV FWHM to account for lifetime and instrument
broadening respectively. The $E_{\rm F}$
corresponds to that of the LDA calculations.}
\end{figure}

In Fig.~\ref{f:rixs}, we compare the RXES spectra at excitation energy (f)
with angle-integrated PES measurements, recorded with a photon energy
of 80~eV.  These PES measurements (Fig.~\ref{f:rixs}a) exhibit the typical
double-peaked structure of the LHB and QP that have previously been reported
for CaVO$_3$ and SrVO$_3$,\cite{eguchi2006,yoshida2010} including the transfer
of spectral weight from the QP to the LHB, and the shift to lower binding
energies of the LHB in CaVO$_3$.  Indeed, such an evolution is predicted by
DMFT (the DMFT of Nekrasov {\em et al.}\cite{nekrasov2005} are reproduced,
multiplied by the Fermi function, for comparison in Fig.~\ref{f:rixs}),
and stems from the increased importance of electron-electron correlations
(often characterised as $U/W$) compared with the narrower bandwidth, $W$,
of CaVO$_3$. In Fig.~\ref{f:rixs}b, the V $3d$ PES spectrum of CaVO$_3$ is
shown after convolution with a Lorentzian of 0.1~eV HWHM (half-width half
maximum)
and a Gaussian of 0.35~eV FWHM to approximate the finite lifetime and instrument
broadening respectively, and shifted in energy to
align the high energy QP features. Whilst the double-peaked structure of the
broadened PES spectrum is reproduced by the bulk-sensitive RXES measurement,
the energy separation between the LHB and QP is larger in the bulk. On the
other hand, a similar broadening of the DMFT CaVO$_3$ results leads to an
overestimation of this separation. These results indicate that the LHB appears
closer to the QP in surface-sensitive measurements, whereas the bulk-sensitive
RXES measurements are in better agreement with DMFT predictions based on
a first-principles calculation of $U$.\cite{nekrasov2005} We note that
in an early PES study surface disorder was found to shift the LHB further
from $E_{\rm F}$,\cite{aiura2001} suggesting the differences we observe
are intrinsically linked to the surface.  Indeed, such an enhancement
in the effects of electron correlation at the surface is predicted by
DMFT calculations of the SrO-terminated layer of SrVO$_3$, in which the
LHB was found to be 40\% closer to the QP than calculations of the
bulk.\cite{ishida2006}
We note that this effect is already present in calculations of the idealized
(unrelaxed) surface,\cite{ishida2006} indicating the enhancement in
correlated behavior is a fundamental property of the surface.
In the lower part of Fig.~\ref{f:rixs}b, RXES spectrum
(f) is shown of the CaVO$_3$ 46~nm thin film, alongside the same (broadened)
PES spectrum (and shifted to approximately align the QP). As can be seen,
the PES and RXES features are in much better agreement for this sample, in
which the surface contribution to the RXES measurement dominates over the
bulk. The longer low-energy tail of this spectrum (extending
below 512~eV) may represent a small bulk-like volume of the thin-film.

Finally, in Fig.~\ref{f:okxes}, the O $K$-edge normal XES spectra of CVO
and SCVO are shown alongside the O and V PDOS from the LDA calculation. The
low-energy shoulder at $\sim 523$~eV is due to the bonding states, and
is relatively more intense for SCVO. This enhanced V-O hybridization
is predicted by the LDA, in which the bonding O PDOS is predicted to be
stronger for SrVO$_3$ compared with CaVO$_3$, and is also reflected by
the stronger V $3d$ -- O $2p$ feature in the RXES measurements of SCVO
(Fig.~\ref{f:rxes}b). The discrepancy in the relative
intensities of the bonding and non-bonding features between experiment and
theory is most likely due to matrix element effects, which
are not considered here.
At higher energies, a substantial shoulder near 529~eV
is observed, and can be attributed to the weak mixing of the O wavefunctions
in the V $3d$ manifold. Additionally, approximately 1.5~eV below, there
is a second poorly-resolved feature, which is slightly more intense for
CVO than SCVO, and which appears at a gap in the LDA DOS.  Tentatively, we
speculate that this may represent emission from the LHB. If so, these spectra
indicate that there is substantial mixing of the O $2p$ states with the LHB,
and the assignment of the incoherent feature may be more complex than the
LHB in its conventional meaning.
Such a scenario has recently been proposed via extended cluster model
calculations.\cite{mossanek2010} Further measurements and calculations are
needed to accurately establish the nature of this feature.

In summary, we have shown that RXES is a sensitive {\em bulk} probe of
correlated electron behavior in Sr$_x$Ca$_{1-x}$VO$_3$. The application
of this technique to other correlated materials, particularly in tandem
with ARPES measurements, may shed new light on the differences between
electron correlations at the surface and in the bulk in systems such as
topological insulators\cite{schmidt2012} and high-$T_c$
superconductors\cite{koralek2006}.
In Sr$_x$Ca$_{1-x}$VO$_3$, we find better agreement in the
energetics of the spectral features between DMFT\cite{nekrasov2005,ishida2006}
and our RXES measurements than we do with PES. Comparison with a
thin-film sample supports our interpretation.

{\bf Acknowledgements}. The authors would like to thank T.\ Balasubramanian for
valuable discussions regarding the experiment.
The Boston University program is supported in part by
the Department of Energy under Grant No.\ DE-FG02-98ER45680. Supported also in
part by the Boston University/University of Warwick collaboration fund.
The Advanced
Light Source, Berkeley, is supported by the US Department of Energy under
Contract No.\ DEAC02-05CH11231. G.B.\ gratefully acknowledges
financial support from EPSRC Grant EP/I007210/1.
M.G., J.W.L.\ and S.A.W.\ gratefully acknowledge
financial support from the Army Research Office through MURI grant No.\
W911-NF-09-1-0398.

\end{document}